# Dynamics of the hysteretic voltage-induced torsional strain in tantalum trisulfide

**J. Nichols, C. Sandamali Weerasooriya, and J.W. Brill**

Department of Physics and Astronomy

University of Kentucky

Lexington, KY 40506-0055, USA

jwbrill@uky.edu

**Abstract.** We have studied how the hysteretic voltage-induced torsional strain, associated with charge-density-wave depinning, in orthorhombic tantalum trisulfide depends on square-wave and triangle-wave voltages of different frequencies and amplitudes.  The strains are measured by placing the sample, with a wire glued to the center as a transducer, in a radio frequency cavity and measuring the modulated response of the cavity.   From the triangle waves, we map out the time dependence of the hysteresis loops, and find that the hysteresis loops broaden for waves with periods less than 30 seconds.  The square-wave response shows that the dynamic response to positive and negative voltages can be quite different.  The overall frequency dependence is relaxational, but with multiple relaxation times which typically decrease with increasing voltage.  The detailed dynamic response is very sample dependent, suggesting that it depends in detail on interactions of the CDW with sample defects.

# I. Introduction

An important advance in nanoelectronics will be the development of simply constructed nano-electromechanical components, such as single component actuators. Recently, Pokrovskii *et al* [1,2] discovered that several quasi-one dimensional charge-density-wave (CDW) conductors twist under application of a dc voltage, with effective piezoelectric coefficients $\sim 10^{-4}$ cm/V, orders of magnitude larger than conventional piezoelectrics. While the initial work was on fiber-like crystals with micron-size transverse dimensions, the twist angle appeared to scale inversely with transverse dimensions [1], giving the hope that large effects would also be manifest in CDW nanocrystals.

Most experiments [1-3] have been on the orthorhombic polytype of tantalum trisulfide (*o*-TaS$_3$), which undergoes a transition into a semiconducting CDW state at $T_c$ = 220 K [4]. Below $T_c$, the CDW can be depinned and made to slide by application of voltages above a threshold, $V_T$ [4], giving rise to a number of novel electronic and mechanical properties [5], including a large (25%) increase in its low-frequency shear compliance [6]. Pokrovskii *et al* found that near $V_T$ the crystal twisted, by typically $\sim 1°$ corresponding to torsional strains $\sim 10^{-4}$, with the direction of twist reversing when the voltage was reversed past -$V_T$, so that the twist angle was a hysteretic function of voltage [1] (e.g. see Figure 1 below). While these hysteretic changes appeared to be quite sluggish [2], they also found that *o*-TaS$_3$ exhibited a much faster and non-hysteretic, but also much smaller, voltage-induced torsional strain (VITS) that was not tied to CDW depinning [2].

In preliminary work [3], we studied the hysteretic VITS signal in crystals of *o*-TaS$_3$ and found that it commenced at voltages below the sliding threshold, suggesting that the torsional crystal strain was caused by deformations of the CDW rather than CDW current. The reason why CDW deformations, usually assumed to vary only along the sample length [7,8], i.e. distance from current contacts, should become chiral remains mysterious, however, as discussed further below. Furthermore, we observed no complimentary "torsional strain induced" voltage when straining the crystal with an applied torque when biased below $V_T$ [3], although the large noise associated with CDW depinning [5] limited our sensitivity. Finally, we started measurements of the dynamics of the hysteretic VITS by studying the oscillating strain in response to square-wave voltages, showing that the response remained frequency dependent at frequencies even below 1 Hz. In the present work, we extend these measurements to give a more complete picture of the dynamics of the hysteretic VITS in *o*-TaS$_3$.

## II. Experimental Details and Results

In previous work [1,3], one end of the $o$-TaS$_3$ crystal was rigidly mounted on an electrical contact while the second contact was made with a very fine wire, leaving that end of the sample free to twist. For our present measurements, we used a much simpler and more robust mounting method. Both ends of the crystal were rigidly mounted on contacts, but a gold film was evaporated on approximately half the sample to electrically short it out and keep the CDW pinned on that side, which would then act as a spring opposing the torsional strain induced on the other half and halving the net voltage-induced torsional strain. As in Reference [3], the sample was placed in a helical resonator RF cavity [9]. A small (~ 1 mm long, 25 μm diameter) steel wire was glued to the center of the sample and placed near the tip of the helix, so that the twisting sample changed the resonant frequency of the cavity. Oscillating strains (e.g. see Figures 3 and 4 below) were measured by measuring the modulated output of the cavity with a lock-in amplifier tuned to the oscillation frequency. The time dependence of strains (e.g. see Figures 1 and 2) were measured by driving the cavity with a frequency modulated carrier tuned slightly off-resonance, and measuring the modulation amplitude with a digital oscilloscope. For both cases, measurements of the strain are relative, but we estimate that the order of magnitude of the twist is ~ 1° [1]. (For the time-dependent measurements, the vertical offsets of the strains are also arbitrary.) Use of the steel wire also allowed us to apply torque to the sample with a magnetic field, which was useful in tuning the electronics as well as measuring the voltage dependence of the shear compliance of the sample [3,6].

Measurements were made on two crystals, referred to as C and D, with dimensions ~ 5 mm x 15 μm x 4 μm, and torsional resonant frequencies ~ 560 Hz (C) and 800 Hz (D). All measurements were at T = 77 K, where the resistances of both samples were ~ 35 kΩ and their threshold voltages, i.e. where the shear compliance began to increase [3,6] were $V_T$ ~ 300 mV (C) and $V_T$ ~ 200 mV (D). Note that these threshold voltages are several times larger than those of samples A and B of Reference [3], which were of similar lengths, indicating that the present samples were slightly less pure [5].

The dynamics of the VITS hysteresis loops, shown in Figure 1, were measured by applying triangle wave voltages of different periods to the samples. Note that the two samples twisted in opposite directions as the voltage was swept from positive to negative and back. For sample D, the twisting commences at voltages below $V_T$ (shown by vertical arrows), but this is less obvious for sample C. While the strain is not a symmetric function of voltage, the lack of symmetry is less striking than for the samples of Ref. [3]. The most prominent non-symmetric feature for sample D is that the magnitude of the strain decreases for increasing positive voltages > $2V_T$, while this does not happen for negative voltages. For sample C, on the other hand, the lack of symmetry seems to manifest itself more in the dynamics, with the torsion changing much more rapidly for positive voltages than for negative. Overall, there does not seem to be a strong period dependence to the magnitude of the VITS, as long as the voltage is swept past $2V_T$, but the voltages at which the strain begins to change and then saturates decrease, i.e. the hysteresis loops "narrow", with increasing period, even for periods as long as T = 33 s; e.g. compare the hysteresis loops for T = 33 s and T = 83 s for sample D.

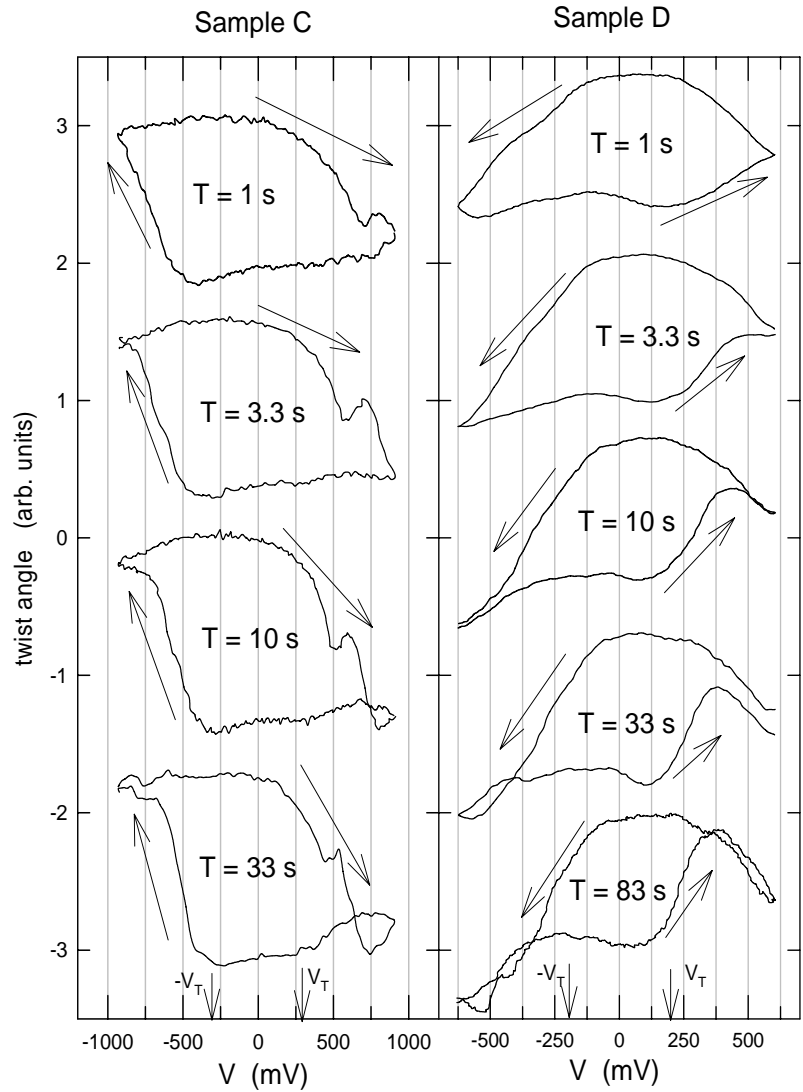

**Figure 1.** Torsional strain (i.e. twist angle) vs. voltage for the two samples when symmetric triangle-wave voltages of different periods (T) are applied to the sample. The vertical offsets are arbitrary. For each sample, the vertical units for each curve are the same, but they may be different for the two samples. Each curve represents the average of up to 100 cycles. The vertical arrows at the bottom show the threshold voltages where the shear modulus begins to soften [3,6].

Application of the VITS effect will presumably involve switching the voltage between positive and negative values. The time dependence of the torsional strain in response to symmetric square wave voltages of different amplitudes and periods is shown in Figure 2. The time dependence of the applied voltage is shown in the lower right panel; again notice that the polarity of the VITS is the opposite for the two samples. For $V = 3V_T$ for both samples, the magnitude of the torsional strain decays with time for long period square waves during one-half cycle (when the strain is "negative"), but not for shorter periods, smaller voltages, or "positive" strains.

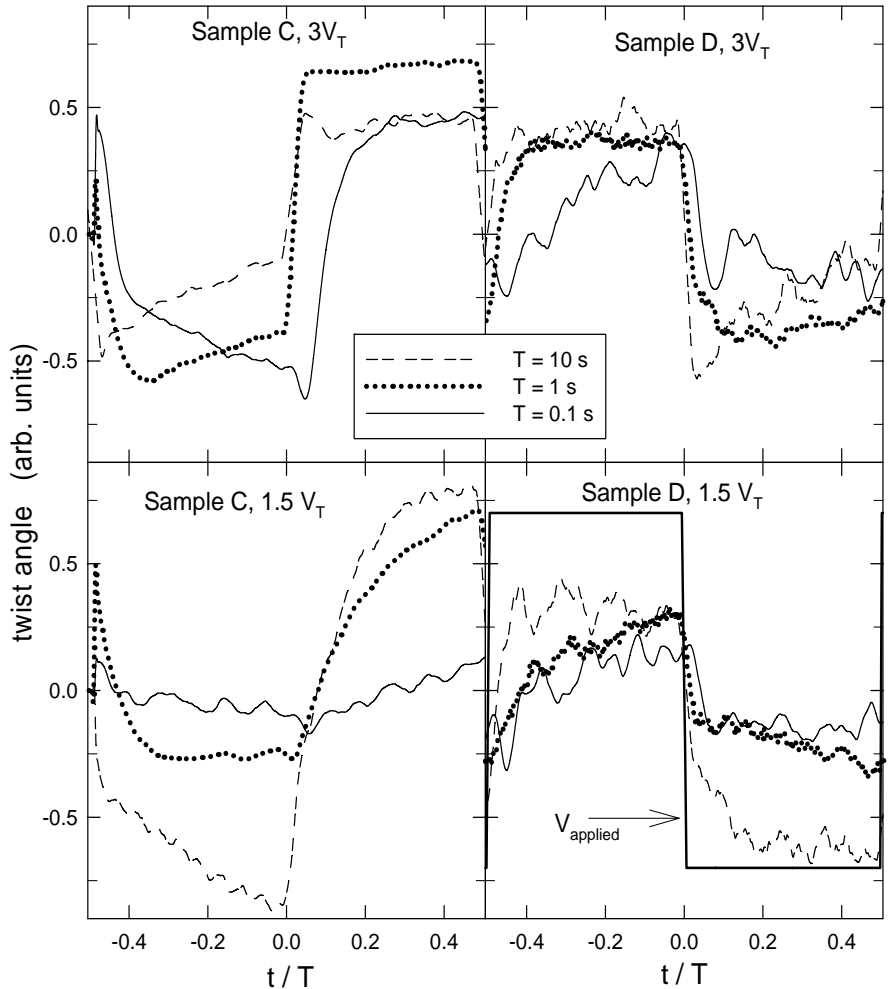

**Figure 2.** Torsional strain vs. time for square-wave voltages of different magnitudes and periods (T) for the two samples. The phase of the applied square wave is shown in the lower right panel. The vertical offsets are arbitrary. For each sample, the vertical units for each curve are the same, but they may be different for the two samples. Each curve represents the average of up to 300 cycles.

Note that for sample C with $T = 10$ s, the magnitude of the oscillating strain is greater at 1.5 $V_T$ than at 3 $V_T$. For sample C, there is again considerable asymmetry between the two half cycles. When the voltage switches from negative to positive, the strain jumps to a positive value $\sim 5$ ms before starting to relax to its negative value. For $V = 1.5 V_T$ and $T = 1$ s, the strain quickly (in $\sim 0.1$ s) saturates in this negative half cycle but does not saturate during the positive half cycle. On the other hand, for $V = 1.5 V_T$ and $T = 10$ s, the strain is still increasing linearly during the whole negative half cycle, while it saturates in $\sim 3$ s during the positive half cycle. These very different behaviors during the negative half cycles for waves of different periods imply a complex dependence of the torsional strain on its history before the voltage switch. While the data for sample D is much noisier, it does not appear to have these large dynamic asymmetries.

To quantify the response to square wave voltages, we measured the voltage dependence of the in-phase and quadrature strains, $\varepsilon_\omega$, with a lock-in amplifier tuned to the fundamental square-wave frequency, $\omega$ (which was kept well-below the sample's torsional resonant frequency so that the response to a fixed oscillating torque would be independent of frequency). The responses at 5 Hz and 0.2 Hz for the two samples are shown in the insets to Figures 3 and 4. The behavior of both samples is intermediate between that of (previously reported) Samples A and B [3]: i) at 5 Hz, the in-phase strain for C and D increases monotonically with voltage while the quadrature response saturates, similar to the behavior of sample A [3]; ii) at 0.2 Hz, the in-phase response of C and D peaks near $2V_T$ while the quadrature response approaches negative values at the highest voltages, corresponding to the long-time decay of the strain noted above and similar to the response of sample B [3]. Note that significant delays in the strain with respect to the applied voltage would cause the in-phase response to become negative, but the $\sim 5$ ms delays observed for sample C are too short to cause such inversions at the measured frequencies.

From a series of measurements similar to these, we determined the frequency dependence of the in-phase and quadrature responses at several voltages; some results are shown in the main panels of Figures 3 and 4. For sample C (Fig. 3), there are two peaks in the quadrature response corresponding to two drops in the in-phase response (most clearly seen at intermediate voltages), suggesting two distinct relaxation processes. Neglecting the long-time decay, the curves in Figure 3 are consequently fit to

$$\varepsilon_\omega = A_1/(1-i\omega\tau_1) + A_2/(1-i\omega\tau_2) \qquad (1)$$

and the voltage dependence of the parameters is shown in Figure 5. For voltages above the hysteresis loop ($V>2V_T$), the time constants decrease very rapidly with voltage, e.g. $\tau_1 \sim V^{-4.5}$ and $\tau_2 \sim V^{-8}$. These are much stronger voltage dependences than observed for the relaxation times associated with the electro-optic response, $\tau_{electro-optic} \sim V^{-1.5}$ [10], where $\tau_{electro-optic}$ was associated with phase-slip enhanced diffusion of longitudinal CDW phase deformations [8]. At smaller voltages, as the sample enters the hysteresis loop and the amplitudes drop, the slower relaxation process dominates.

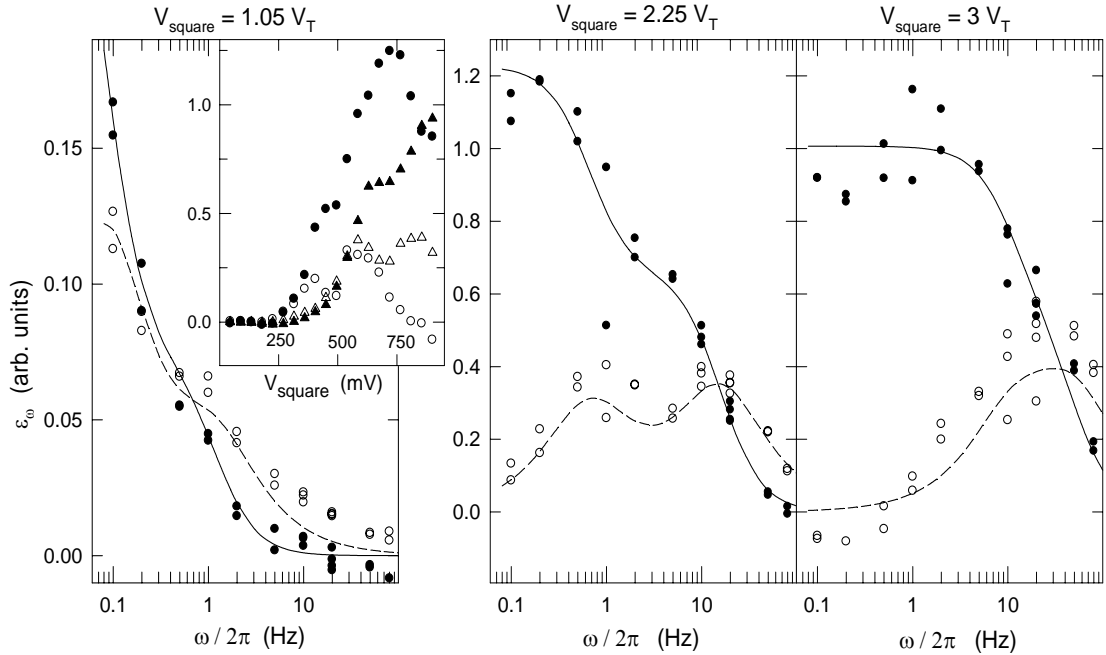

**Figure 3.** Torsional strain at the fundamental frequency for square-wave voltages as functions of frequency at a few voltages for sample C. Filled symbols show the response in phase with the applied square wave and open symbols the quadrature response. The curves show the fits to Eqtn. (1). The vertical units are the same in each panel. Inset: The voltage dependence of $\varepsilon_\omega$ for 5 Hz (triangles) and 0.2 Hz (circles) square-waves; the filled symbols are the in-phase response and the open symbols the quadrature response.

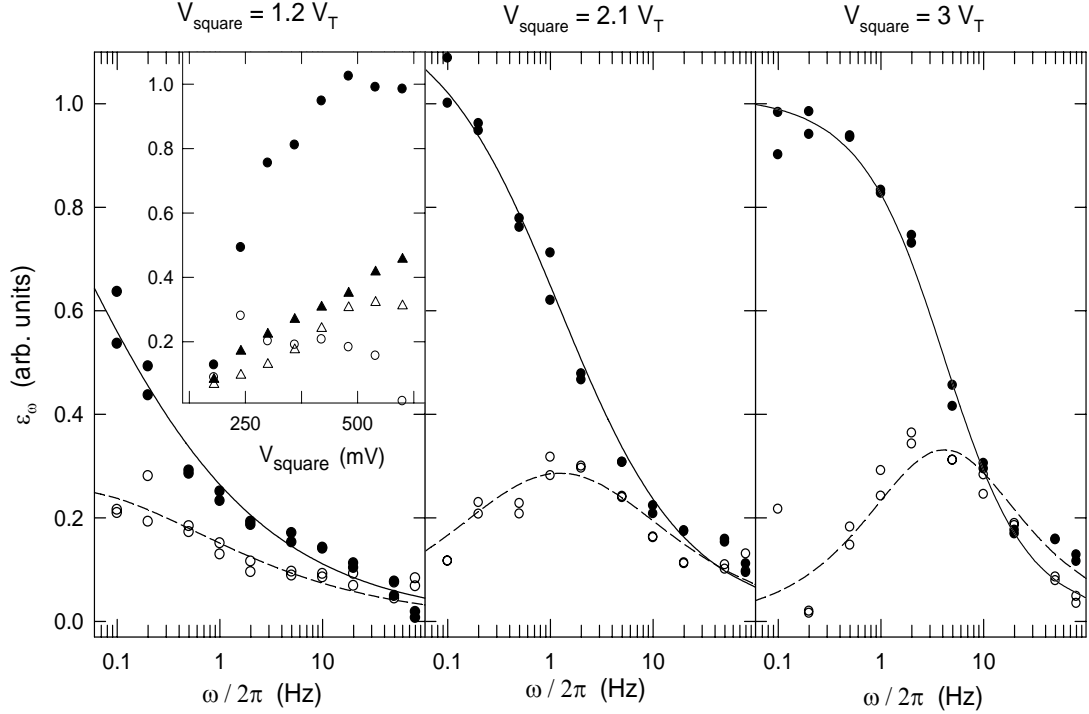

**Figure 4.** Torsional strain at the fundamental frequency for square-wave voltages as functions of frequency at a few voltages for sample D. Filled symbols show the response in phase with the applied square wave and open symbols the quadrature response. The curves show the fits to Eqtn. (2). The vertical units are the same in each panel. Inset: The voltage dependence of $\varepsilon_\omega$ for 5 Hz (triangles) and 0.2 Hz (circles) square-waves; the filled symbols are the in-phase response and the open symbols the quadrature response.

Figure 4 shows the frequency dependence at a few voltages for Sample D, for which there is only a single drop in the in-phase response and peak in the quadrature response. However, the magnitude of the quadrature peak is less than half of the change in the in-phase response, indicating that one needs to include a distribution of relaxation times to fit the response, i.e. $\varepsilon_\omega = \int a(\tau) \, d\ln\tau / (1-i\omega\tau)$. For concreteness, we fit the strain to the Cole-Cole expression [11,12]

$$\varepsilon_\omega = A_0/[1+ (-i\omega\tau_0)^\gamma] \quad (2),$$

for which the distribution of relaxation times is given by [12]

$$a(\tau) = (A_0/\pi)(\tau/\tau_0)^\gamma \sin(\gamma\pi) / [1 + 2(\tau/\tau_0)^\gamma \cos(\gamma\pi) + (\tau/\tau_0)^{2\gamma}]. \quad (3)$$

The parameters of the fits are shown in Figure 5. Also shown in the bottom panel are the widths, $\Delta \ln(\tau/\tau_0)$ (full width half maxima), of the distributions in $\ln(\tau)$ implied by the values of $\gamma$ and Eqtn. (3).

At the lowest voltages, as the sample enters the hysteresis loop, the relaxation peaks fall outside our measured frequency range and the values of the average relaxation time become very uncertain, but for $V \geq 1.5V_T$, the values of $\tau_0$ become well defined. At the highest voltages, sample D is distinctly slower than sample C, but at intermediate voltages, the values of $\tau_0$ lie between those of $\tau_1$ and $\tau_2$. Note that in this region the time constant distribution of sample D is one-to-two decades wide. $\tau_0$ also has a much weaker voltage dependence than $\tau_1$ and $\tau_2$. While the voltage dependence of $\tau_0$ is similar to that of the electro-optic relaxation time, $\tau_0$ is much larger than $\tau_{electro-optic}$ (< 2 ms [10]).

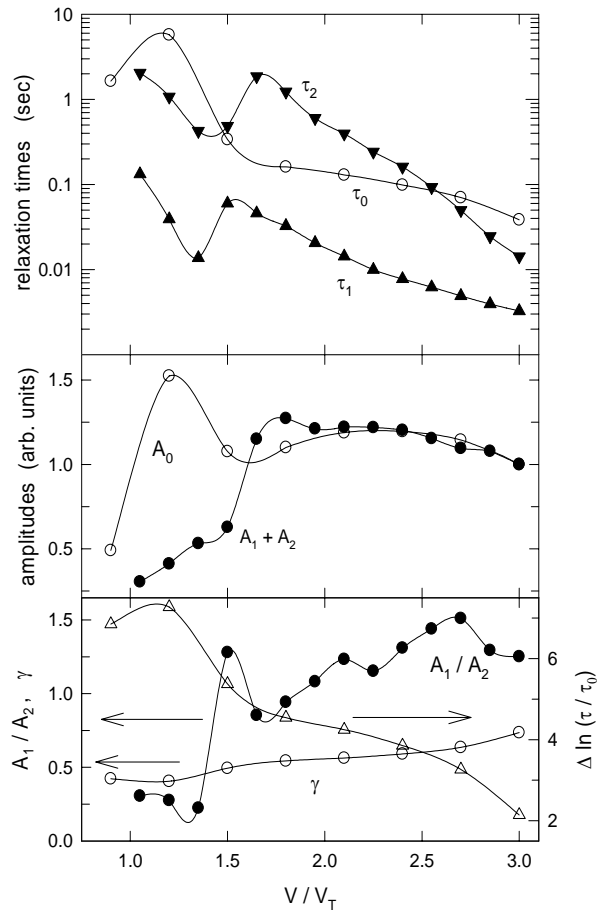

**Figure 5.** Voltage dependence of the fitting parameters to Eqtn. (1) for sample C (filled symbols) and Eqtn. (2) for sample D (open symbols). Also shown in the bottom panel are the widths of the distribution in relaxation times for sample D. The curves are guides to the eye.

## III. Discussion

As mentioned above, the fact that the onset voltage of the hysteresis loops is typically below the sliding threshold suggested that the VITS is caused by CDW deformations (i.e. phase gradients) rather than CDW current [3]. Such is also suggested by the form of the VITS hysteresis loops, i.e. after applying voltage to strain the crystal, the strain remains when the voltage and CDW current are removed, because such persistence is also true of (longitudinal) changes in the CDW phase [13]. We therefore assume that the VITS is caused by azimuthal deformations of the CDW phase, presumably near the mechanically free contact, coupling to the lattice. If the coupling of the phase gradients to crystal strain is anisotropic, as expected, such gradients, even though not chiral themselves, can lead to torque on the sample.

Note that coupling of longitudinal strains and CDW deformations, created by temperature changes, have been reported [14], and that Hoen *et al* [15] have shown that the sample *length* could be a hysteretic function of voltage as a result of such coupling when the strain response was relaxational. Hoen *et al's* model was motivated by their observation of small, sluggish hysteresis loops in the sample length ($\Delta L/L \sim 10^{-6}$ with time constants ~ 10 s), very similar in shape to the VITS loops [15]. (Since a change in length requires a non-zero average sample strain while voltage induced CDW phase changes have opposite signs at the two sample contacts, the length change was interpreted as a finite size effect [15].)

As an example of crystal strain hysteresis loops that could result from voltage-induced CDW deformations, we have simulated the voltage dependence of the strain ($\varepsilon$), assuming that $d\varepsilon/dt = (\varepsilon_0(V)-\varepsilon)/\tau$; we assumed that the "equilibrium" strain $\varepsilon_0(V)$ grows with voltage above an onset $V_{ON}$ and a relaxation time which drops at high voltages [7,10,13]. In particular, we took $\tau = \infty$ for $V < V_{ON}$, $\tau = \tau_0$ and $\varepsilon_0(V) = (V-V_{ON})/(V_{SAT} - V_{ON})$ for $V_{ON} < V < V_{SAT}$ (the voltage at which the strain stops growing [3]), and, $\tau = \tau_0(V_{SAT}/V)^6$ (as observed for sample C) and $\varepsilon_0(V) = 1$ for $V>V_{SAT}$, with corresponding negative values of $\varepsilon_0$ and similar time constants for negative voltages. Results, for $V_{SAT} = 1.5\ V_{ON}$ and different period triangle waves are shown in Figure 6. The results capture the general shapes of the hysteresis loops, in particular how slowly the loops narrow with increasing period, even for periods much larger than $\tau_0$.

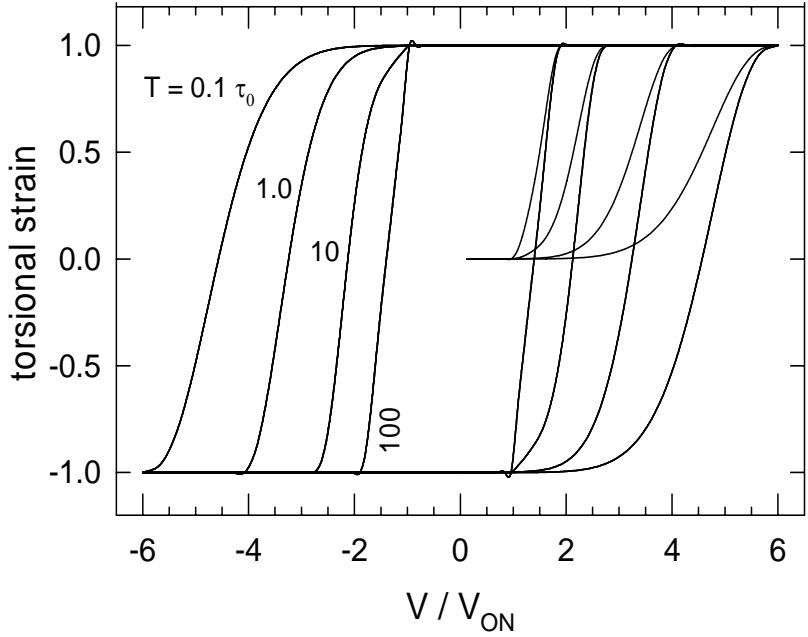

**Figure 6.** Simulated torsional strain hysteresis loops for triangle waves with amplitude $6V_{ON}$ for different periods T, as described in the text.

The outstanding questions are therefore i) what causes such long relaxation times and ii) what causes the azimuthal CDW deformations in the first place. For longitudinal phase CDW deformations, the time constant near threshold is determined by diffusion of the CDW phase along the length of the sample, i.e. $\tau_0 \propto L^2/D$, while at higher voltages phase-slip, associated with the CDW current, allows the phase to change more quickly [8]. The CDW phase diffusion constant is expected to be proportional to the phason velocity [8, 16], and excellent agreement was found for longitudinal phase deformations of the CDW in quasi-one dimensional blue bronze ($K_{0.3}MoO_3$) measured electro-optically [16]. For $o$-TaS$_3$, the low-voltage VITS time constant is two orders of magnitude larger than $\tau_{electro-optic}$ [10]. Since the sample width is also two orders of magnitude less than the length, if the VITS dynamic response was similarly diffusive, it would imply a transverse diffusion constant six orders of magnitude less than the longitudinal diffusion constant. While we know of no estimates of the anisotropy of the phason dispersion, it seems unlikely to be so large. Instead, we suggest that changes in the chiral deformations of the CDW that lead to the VITS are hindered by the interaction of the deformations with crystal defects, and the strong sample dependence of the dynamic response of the present samples and those studied in Ref. [3] certainly suggests that different types of crystal defects can play important roles.

Finally, we come to the question of what may initially cause azimuthal CDW deformations. No chiral feature has been observed in either the crystal or CDW structure [1]. Note that if, instead of being normal to the sample length as usually assumed, the CDW wavefronts are tilted or crooked on one surface but not the opposite surface, this could give rise to azimuthal gradients of the CDW phase which can lead to twisting of the sample. The CDW undergoes compressions and expansions at the current contacts with voltage reversals which would then lead to an oscillating twist. While the distortion of the wavefronts may result from crystal defects, an interesting possibility is that it arises from the contact itself, which may not be perfectly straight or normal to the sample axis. The resulting slanted or crooked equipotential lines may cause shear stress of the CDW on the surface, and the resulting azimuthal CDW phase gradients would only heal in the longitudinal phase coherence length, $\sim$ 10 - 100 μm [17]. Experiments to investigate whether tilted contacts lead to torque are underway.

In conclusion, our present results, together with those of Ref. [3] show that while there are common features about the hysteretic VITS dynamics, details are sample dependent. In common is that the effect is essentially relaxational, with an average relaxation time that decreases with voltage and has a value of $\sim$ 1 sec near threshold. However, the voltage dependence is sample dependent, as is the detailed behavior after reversing the voltage; in some cases, the VITS is observed to decay at long times and in some cases jumps and delays are observed before the strain reverses. These effects are also polarity dependent. The long time constants suggest that the hysteretic VITS results from interactions of deformations of the CDW with lattice defects, and the sample dependent dynamics suggest that different types of defects may affect the strain differently.

One of us (JWB) acknowledges the inspiration provided by his long-time colleague, Peter Eklund, for using novel probes to study unusual properties of materials. We also thank V. Ya. Pokrovskii and D. Dominko for helpful suggestions and R.E. Thorne for providing crystals. This work was supported by the U.S. National Science Foundation under Grants No. DMR-0800367, EPS-0814194, and DMR-0400938.